\documentclass[usenatbib,useAMS]{mn2e}

\usepackage{epsf}
\usepackage[dvips]{epsfig}
\usepackage{times}

\begin{document}

\label{firstpage}

\title[Mass-Loss vs.\ Dynamical Friction]{The Influence of Mass-Loss
  from a Star Cluster on its Dynamical Friction -- I. Clusters without
  Internal Evolution} 

\author[Fellhauer \& Lin]{M. Fellhauer$^{1,2}$
  \thanks{madf@ast.cam.ac.uk}, D.N.C. Lin$^{2}$
  \thanks{lin@ucolick.org} \\
  $^{1}$ Institute of Astronomy, University of Cambridge, Madingley Road, 
  Cambridge CB3 0HA, UK\\
  $^{2}$ UCO/Lick Observatory, UC Santa Cruz, 1156 High Street, Santa
  Cruz, CA 95064, USA}
 
\pagerange{\pageref{firstpage}--\pageref{lastpage}} \pubyear{2006}

\maketitle

\begin{abstract}
  Many Local Group dwarf spheroidal galaxies are found in the Galactic
  halo along great circles in the sky.  Some of these stellar systems
  are thought to be the fragments of larger parent galaxies which have
  once intruded into and were torn apart by the tide of the Galaxy.
  Supporting evidences for tidal disruption are found in the form of
  stellar tidal bridges and tails along the orbits of some dwarf
  galaxies and globular clusters.  In this study, we investigate the
  influence of mass-loss from star clusters or dwarf galaxies on the
  rate of their orbit decay due to the effect of dynamical friction.
  Using a series of numerical N-body simulations, we show that stars,
  which become unbound from their host-systems, but remain in their
  vicinity and share their orbits, still contribute to the mass
  responsible for the dynamical friction.  As a rule-by-thumb, the
  magnitude of dynamical friction at any instance can be approximated
  by the bound mass plus half of the mass which has already become
  unbound during the proceeding Galactic orbit.  Based on these
  results, we suggest the tidal disruption of relatively massive
  satellite stellar systems may be more abrupt than previously
  estimated.
\end{abstract}


\section{Introduction}
\label{sec:intro}

In the widely adopted $\Lambda$CDM scenario, large normal galaxies are
formed through the mergers of much smaller entities, dwarf galaxies
\citep{blu84,nav95}.  In the halo of the Galaxy, hundred of such
entities are anticipated \citep{moo99}, far exceed the previously
known satellite stellar systems.  In recent years, however, many new
satellite dwarf spheroidal galaxies are being discovered, namely Ursa
Major \citep[UMa;][]{wil05}, Canes Venatici \citep[CanVen;][]{zuc06a},
Bootes \citep[Boo;][]{bel06b} and Ursa Major II \citep[UMa
II;][]{zuc06b}.  The study of their dynamical evolution and star
formation history has becoming an important aspect in the
determination of poorly known halo structure and the development of
galaxy formation theory.

As they orbit around the halo of their host galaxies, satellite
galaxies encounter the effect of dynamical friction and undergo
orbital decay.  When they migrate toward the centre of their host
galaxy, the increasingly intense background tidal perturbation leads
to the disruption of loosely bound satellites.  In the vicinity of
several satellite galaxies, streams of escaping stars have been
discovered \citep[e.g.][]{bel06a}.  Similar tidal debris is also
found along the orbits of some loosely bound globular clusters
\citep{ode03,bel06c}.

As a theoretical construct, the dynamical-friction process was first
described by \citet{cha43}.  Under the idealised assumption that an
undisruptable stellar system moving through a homogeneous background
of stars (with a Maxwellian phase space distribution and an isotropic
velocity dispersion), the magnitude of dynamical friction is
proportional to the mass of the system.  But, for disintegrating
stellar systems with a declining mass within their tidal radii, it
remains uncertain whether this mass includes the contribution from
those stars or gas which became gravitationally detached during the
most recent passages.  In principle, their gravity continues to
exchange momentum with the background constituents. However, their
weakened gravitational coupling with their originally parent galaxies
also reduces the rate of momentum transfer between them.

The efficiency and outcome of this mass-loss process are sensitively
determined by the nature and magnitude of the dynamical friction.  If
the orbital decay rate is directly proportional to the mass of the
residual stars which remain gravitationally bound to the sinking and
disrupting satellite stellar systems, their cores' migration would be
quenched and structure would be preserved as their outlying stars
become detached. In an attempt to account for its stellar metallicity
dispersion, it has been proposed that the globular cluster
$\omega$-Centauri may indeed be the core-remnant of a destroyed
satellite dwarf spheroidal galaxy \citep{tsu04,bek03,hil00}.  If,
however, the relative rapid orbital decay rate is sustained despite
the tidal disruption of the satellite dwarf galaxies or globular
clusters, they would completely disrupt and their constituent stars
would add to the stellar components of the outer halo in form of
streams \citep{iba02,mar04}.

Dynamical friction is a process which also plays a crucial role in
other stellar-dynamical problems.  Recent observations of the Galactic
Centre, for example, revealed very young star clusters, with ages of a
few Myr, inside the inner parsecs of our Galaxy
\citep{tam93,kra95,ger01}.  Because this location is believed to be
too hostile to form star clusters, these objects must have formed
further out and have sunk to the actual position within their short
lifetime.  The mechanism to shrink their orbit is dynamical friction
\citep{cha43}.  Once again, whether the rate of orbital decay is
affected by the tidal disruption of the clusters determines the fate
of these stellar systems.

The above discussions indicate that theoretical and numerical studies
to establish a better understanding of dynamical friction are
important tasks. Recent numerical studies on this topic focused on the
influence of resolution of the used N-body codes on the results of
dynamical friction \citep{spi03}, the influence of the density profile
on the Coulomb logarithm \citep{pen04,jus05} or the influence of the
mass-loss on the orbital decay \citep{mac02,fuj06}.

In this paper, we investigate the interplay between dynamical
friction, which leads to the sinking of the object, and the mass-loss
due to tidal heating by means of numerical N-body simulations.  The
overall objectives of this investigation are similar to those
motivated the investigation by \citet{fuj06}.  But, in our
simulations, we follow the evolution of the satellite further, beyond
the stage of complete disruption.  We also take into account the
artificial effects of resolution limitations of N-body codes and the
variation of $\ln \Lambda$ with distance to the centre of the
background.  In the next section we briefly recapitulate the standard
theory of dynamical friction followed by the setup of our N-body
models.  In section~\ref{sec:results} we present the results of our
calculations in a systematic manner.  According to the conventional
formula for dynamical friction, the rate of orbital evolution is
determined by the product of the effective mass ($M_{\rm cl}$) of the
satellite and a parameter ($\ln \Lambda$) which measures the spacial
extent of the halo region which effectively responds to the
satellite's gravity.  In order to disentangle these two effects, we
first carry out an idealised simulation on the orbital evolution of a
cluster with a point-mass potential and a constant mass.  We show that
the magnitude of $\ln \Lambda$ artificially depends on the numerical
resolution for close encounters in this idealised model.  However,
this particular model over-estimated the importance of close
encounters. In order to consider a potential with a more realistic
mass distribution, we also carry out a simulation for a cluster based
on a Plummer model.  We consider the case of a compact cluster so that
its mass is preserved over several orbital periods.  We show that due
to the ``soften'' nature of the cluster's potential, the magnitude of
$\ln \Lambda$ is slightly below that deduced for the cluster with a
point-mass potential \citep[see also][]{spi03}.  Together, these
simulations demonstrate the validity for the numerical method in its
ability to resolve the gravity over close stellar encounters.

We present evidences that the unbound, recently torn stars along the
tidal tails also contribute to the dynamical friction on the
cluster's orbital evolution.  We present the simulation of a loosely
bound cluster which is undergoing total disruption.  With the
simulation of this disintegrating cluster and another idealised model
for an unbound moving cluster, we present evidence that the unbound but
nearly co-moving stars contribute to the orbital decay of the residual
core.  We summarise our results and discuss their implications in
section~\ref{sec:conclus}.

\section{Dynamical Friction}
\label{sec:dynfric}

The mechanism to shrink the orbits of star clusters or dwarf galaxies
in the tidal field of their host-galaxy is dynamical friction, which
was studied first theoretically by \citet{cha43}.  In his formula
\begin{eqnarray}
  \label{eq:dynfric}
  \frac{d{\bf v}_{M}} {dt}\! & \! = \! & \! - \frac{4 \pi \ln \Lambda
    G^{2} \rho 
    M_{\rm cl}} {v_{M}^{3}} \left[ {\rm erf}(X) -
    \frac{2X}{\sqrt{\pi}} \exp(-X^{2}) \right] {\bf v}_{M} \nonumber
  \\
  & & 
\end{eqnarray}
for $M_{\rm cl} \gg m$ where $X=v_{M}/\sqrt{2}\sigma$, $v_{M}$ is the
velocity of the object with mass $M_{\rm cl}$, $\rho$ is the mass
density of the surrounding material, $m$ is the typical single mass
and $\sigma$ the typical velocity dispersion of the surrounding
medium, $\Lambda$ describes the ratio of the largest to the smallest
impact parameter:
\begin{eqnarray}
  \label{eq:lambda}
  \Lambda & = & \frac{b_{\rm max} v_{M}^{2}} {G (M_{\rm cl}+m)} \
  = \ \frac{b_{\rm max}}{b_{\rm min,theo}}. 
\end{eqnarray}
This parameter varies with the different environments and the
determination of $\ln \Lambda$ is a crucial point in determining the
sinking time scales.  Even though Chandrasekhar's formula is only a
linear approximation and derived for an infinite, homogeneous
background, numerical studies proved that the formula holds also for
'realistic' backgrounds \citep{bin87,has03}.

We determine $\ln \Lambda$ from the sinking rate in a background
medium in which the mass distribution follows a power law profile 
\begin{eqnarray}
  \label{eq:mass}
  M_{\rm back}(D) = A \cdot D^{\alpha},
\end{eqnarray}
where $D$ is the distance from the centre of the background medium.
For this mass distribution, we can determine the evolution of a
satellite's orbit with the analytical formula derived by
\citet{mac02}:
\begin{eqnarray}
  \label{eq:fitrad}
  D(t) & = & D_{0} \left[ 1 - \frac{\alpha (\alpha+3)} {\alpha+1}
    \sqrt{ \frac{G} {AD_{0}^{\alpha+3} } } \right. \nonumber \times 
  \\ 
  & &  \left. \left( {\rm erf}(X) -
      \frac{2X}{\sqrt{\pi}} \exp(-X^{2}) \right) M_{\rm cl} \ln
    \Lambda t \right]^{\frac{2} {3+\alpha}}.
\end{eqnarray}
Here we denote $D(t)$ as the satellite's distance from the centre of
the background medium.  We utilise this equation below to deduce the
magnitude of $(M_{\rm cl} \ln \Lambda)$ from the value of $D(t=0) =
D_{0}$ and $D(t)$.

In a recent numerical study, \citet{spi03} investigated the in-fall of
a point-mass into the Galactic centre and compared the results of
different numerical schemes.  In this point-mass case the sinking of
the massive particle was limited by the resolution of the used N-body
codes, which in the case of a particle-mesh code is given by the
length of one cell $l$:
\begin{eqnarray}
  \label{eq:bmin}
  b_{\rm min} & \approx & b_{\rm min,theo} + l
\end{eqnarray}
By fitting their results with varying resolutions they determined that
the maximum impact parameter is approximately half the distance to the
Galactic Centre ($a' \approx 0.5$):
\begin{eqnarray}
  \label{eq:bmax}
  b_{\rm max} & = & a' \cdot D
\end{eqnarray} 

Another important outstanding issue concerning the Chandrasekhar's
formula for the dynamical-friction efficiency in Eq.~\ref{eq:dynfric}
is the specification of the mass $M_{\rm cl}$ of the sinking stellar
system.  In contrast to a sinking massive particle with fixed mass
$M$, a star cluster experiences tidal heating and internal evolution,
leading to a significant mass-loss during its life-time until it is
totally disrupted.  For such stellar systems, the constant mass in
Eq.~\ref{eq:dynfric} has to be replaced by a time-dependent
expression.  But stars which become unbound to their parent clusters
do not disperse immediately.  They continue to travel on almost
identical orbits as the parent cluster itself and disperse only slowly
with time.  The question we want to address in this project is whether
these stars are still taking part in the dynamical friction of the
whole cluster and to which extend or if only the bound particles
contribute to the sinking of the star cluster.

\section{Numerical methods and model parameters}
\label{sec:setup}

As prescription for the background galaxy we adopt the model of a
lowered isothermal sphere \citep[e.g.][]{bin87}.  For computational
convenience, we adopt normalised computer units in which the mass
distribution is cut off at $D_{\rm cut} = 10.0$.  We also set the
total mass of the background galaxy to be $M_{\rm tot} = 1000.0$.
Finally, we specify $G=1$ in model units which implies a theoretical
constant circular velocity of $V_{0}=10.0$ and a characteristic
dynamical time $T_{\rm cr}= 1.414$.  The background is represented
with such a phase space distribution among 5,000,000 particles.

The star cluster is modelled as a Plummer sphere with $M_{\rm pl} =
1.0$ and a Plummer radius $r_{\rm pl}$.  The density
distribution of the Plummer sphere is truncated at $r_{\rm cut} = 5
r_{\rm pl}$.  In order to enhance the mass resolution of the cluster,
we use 1,000,000 particles to represent the stars in the cluster
models.  Even though the adopted mass-ratio ($10^{-3}$) between the
satellite and background is more appropriate for simulating the
interaction between dwarf galaxies and their host galaxies, we use the
generic term 'star cluster' or 'cluster' for our object throughout the
text.  An overview of the cluster parameters used in the different
sections of this manuscript is given in Tab.~\ref{tab:param}.

For the numerical simulations, we use a particle-mesh code named {\sc
  Superbox} \citep{fel00} to compute the gravitational interaction
between particles.  This scheme has two levels of higher resolution
sub-grids focusing on the centres of the simulated objects (here
background and star cluster).  During the simulation, stars which are
attached to the star cluster travel within the medium-resolution grid
of the background with a cell-length of $l=0.08$.  The innermost grid
with the highest resolution covers the central area of the background
until $D=1$ with a resolution of $0.016$.  The time-step of the
simulation is adjusted that particles with typical velocities do not
travel more than one grid-cell per time-step which gives a time-step
of ${\rm dt}=0.001$.  Particle-mesh codes are designed to
suppress the effect of two-body relaxation in galaxy simulations
because their characteristic time scales are longer than a Hubble
time.  Therefore, despite the limited number of representative
particles, a star cluster simulated with {\sc Superbox} do not
artificially generate any internal dynamical evolution such as mass
segregation or core collapse.

The background model is first set-up and integrated forward for
several dynamical times to have the numerical particle distribution
adjust to the used grid-code with its resolutions.  This initial
relaxation alters the shape of the mass-profile of the background
mainly in its innermost central part and in the outer envelope.  After
this initial adjustment, we determine the profile and the half-mass
radius.  Around the half-mass radius, the profile continues to be
nearly isothermal.

The presence of the stellar cluster is introduced to the background
particles by inserting an analytical Plummer-potential with growing
mass and a prescribed Plummer-radius according to the subsequently
used star cluster model.  The initial orbit of the cluster is assumed
to be circular with a radius equal to the half-mass radius of
the background.  This computational procedure is carried out in order
for the particles representing the background to adjust adiabatically
to the later inserted star cluster and to avoid spurious reactions of
the background when the star cluster is inserted.  In reality, such
adjustment is expected since most satellite stellar systems were
probably not formed {\it in situ}.  In all our simulations, we also
ensure a similar adiabatic adjustment for the stellar distribution
within the clusters.  The model of the star cluster is also first
simulated separately to have its internal particles to adjust to the
numerical grid-structure.  Finally the ``live'' star cluster is
inserted in exchange to the analytical Plummer-potential at the
half-mass radius with the actual measured circular velocity.

\begin{table}
  \centering
  \begin{tabular}{lllll}
    model & $D_{0}$ & $r_{\rm pl}$ & $r_{\rm t,ini}$ & orbit \\ \hline
    4.1 & 2.78 & point mass & --- & circular \\
    4.2 & 2.78 & 0.05 & 0.25 & circular \\
    4.3 & 2.78 & 0.10 & 0.25 & circular \\
    4.4 & 2.78 & 0.05 & (0) & circular \\
    4.5a & 2.78 & point mass & --- & eccentric $\epsilon=0.4$ \\
    4.5b & 2.78 & 0.05 & 0.25 & eccentric $\epsilon=0.4$ \\ \hline
  \end{tabular}
  \caption{Model parameters of our simulations. First column gives the
  section in the manuscript in which the simulation and its results
  are explained, $D_{0}$ is the initial distance to the centre of the
  background, $r_{\rm pl}$ is the initial Plummer radius with the
  initial cut-off radius being $5r_{\rm pl}$.  $r_{\rm t,ini}$ is the
  initial tidal radius, which is not applicable to the point mass
  simulations and the simulation 4.4 where the self-gravity is
  artificially switched off.  Last column gives the adopted orbit. In
  the eccentric orbit case $D_{0}$ is the initial apo-centre of the
  simulation.  The initial mass in all simulations is 1.0 and the
  cluster is modelled with 1,000,000 particles in all cases except the
  point mass runs.}
  \label{tab:param}
\end{table}

\section{Computational results}
\label{sec:results}

\subsection{The orbital evolution of a cluster with a constant
  point-mass potential}
\label{sec:i19}

\begin{figure}
  \begin{center}
    \epsfxsize=07.5cm 
    \epsfysize=07.5cm 
    \epsffile{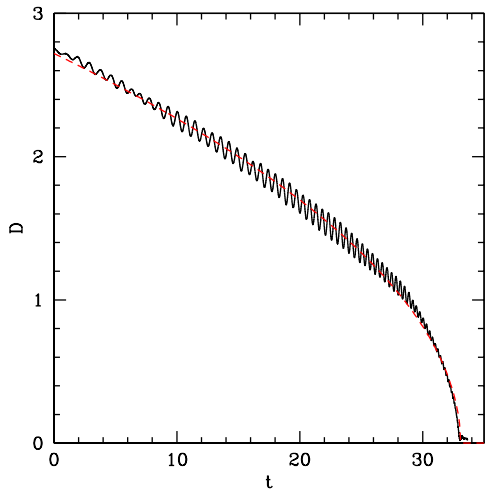}
%%%    \epsffile{i19-rd.eps}
    \caption{The orbital decay of a cluster modelled as a point-mass
      potential.  Shown is the distance $D$ vs.\ the time.  The dashed
      line (red on-line) shows the fitting curve using a single $\ln
      \Lambda$ according to Eq.~\ref{eq:fitrad2}} 
    \label{fig:i19-rd}
  \end{center}
\end{figure}

\begin{figure}
  \begin{center}
    \epsfxsize=07.5cm 
    \epsfysize=07.5cm 
    \epsffile{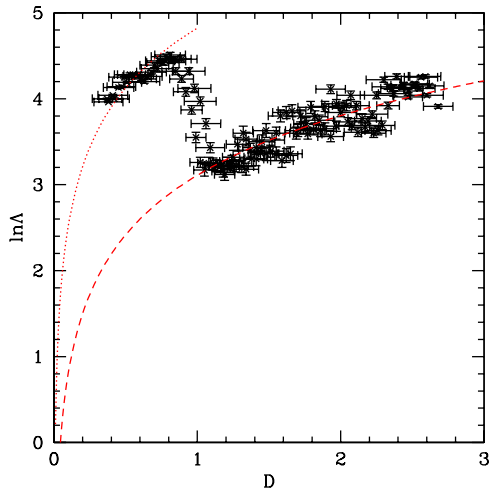}
%%%    \epsffile{i19-fit.eps}
    \caption{Fitting values of $\ln \Lambda$ for each 5 periods of
      oscillations against mean distance $D$.  Lines (red on-line) are
      the fitting curves for the different grid resolutions (dashed:
      medium resolution; dotted: high resolution) as described in the
      main text.(see Eq.~\ref{eq:fitln1}-\ref{eq:fitln3})} 
    \label{fig:i19-fit}
  \end{center}
\end{figure}

In order to disentangle the effect of a changing $\ln \Lambda$ versus
that of a changing mass $M_{\rm cl}$ in Chandrasekhar's formula, we
first simulate the orbital decay of a cluster with a point-mass
potential and a total mass $M_{\rm pm}= 1.0$.  This simulation
provides a determination on the largest value of $\ln \Lambda$.  With
the specified resolution of our code, this determination is possible
because the simulated dynamical friction efficiency is resolution
limited for a point-mass potential whereas $b_{\rm min}$ is
proportional to the characteristic length scales of extended satellite
systems which is generally larger than the resolution.

The star cluster is inserted at the half-mass radius which in our case
is $D_{0} = 2.78$. The characteristic orbital period at the half
mass radius is $\approx 1.75$.  Fitting a power law of the form of
Eq.~\ref{eq:mass} in the range $D \in [1.0:3.5]$ to the background
gives $A=189.9 \pm 0.6$ and $\alpha = 0.935 \pm 0.003$.  Inserting
these values in Eq.~\ref{eq:fitrad} and applying that the velocity of
the star cluster is exactly the measured circular velocity at this
radius, leads to the following theoretical expression for the orbital
evolution,
\begin{eqnarray}
  \label{eq:fitrad2}
  D(t) & = & D_{0} \left( 1.0 - \frac{0.059}{D_{0}^{1.97}} M_{\rm
      cl} \ln \Lambda t \right)^{0.51} .
\end{eqnarray}

In the above expression, $D(t)$ is a function of ($M_{\rm cl} \ln
\Lambda$). We can break this degeneracy with our idealised model in
which the cluster is assumed to have a constant ($M_{\rm cl}=1$ at all
times).  We consider $D_{0}$ as a fitting
parameter because even with the initial orbit of the star cluster
specified to match the circular velocity of the background, small
amplitude oscillation in $D$ during its orbital decay occur.  This
oscillation is due to the discreteness of the time-step and the
'granularity' of the background.  We reduced these numerical
oscillations as much as possible but small and slowly growing
epicyclic oscillations around the expected analytical orbits (Theis,
private communication) remain. This numerical artifact does not affect
our overall results.

Using the implementation of the nonlinear least-squares (NLLS)
Marquardt-Levenberg algorithm in {\it gnuplot}, we find that the
orbital evolution obtained from the numerical simulation (as shown in
Fig.~\ref{fig:i19-rd}) can be fitted with the analytic expression from
Eq.~\ref{eq:fitrad2} with
\begin{eqnarray}
  \label{eq:fit-i19-all}
  D_{0} & = & 2.7174 \pm 0.0003 \\
  \ln \Lambda & = & 3.655 \pm 0.001.
\end{eqnarray}
This result is in good agreement with the previous study of
\citet{spi03} with respect to the dependence of $b_{\rm min}$ and
therefore $\ln \Lambda$ on the resolution of the code (see
Eq.~\ref{eq:bmin}).  The values of $b_{\rm max} = a' \cdot D_{0}$
however differ from the findings in the previous study.  While
\citet{spi03} found values of $a'$ in the range of $0.4$ to $0.6$ our
result implies $a' \approx 1.2$.  A possible reason is that
\citet{spi03} left $b_{\rm min,theo}$ as a free parameter to fit their
results and deduced a value which is three times lower than expected
by Eq.~\ref{eq:lambda}.  Furthermore in both studies the background
was not perfect isothermal introducing a slight dependence of $b_{\rm
  min,theo}$ with distance. 

In order to examine if $\ln \Lambda$ evolves with distance to the
centre of the background, we now take only values from one maximum
(minimum) of the curve to the 5th next maximum (minimum) and again fit
Eq.~\ref{eq:fitrad2} to all the time-slices available.  The measured
values of $\ln \Lambda$ are plotted in Fig.\ref{fig:i19-fit} against
their mean distance $D$ during the used time interval.  The values of
$\ln \Lambda$ clearly decrease with the distance to the centre.

In Section~\ref{sec:dynfric}, we discussed the dependence of
$\ln \Lambda$ on 
the numerical resolution for a cluster with a point-mass potential
(see Eq.~\ref{eq:bmin}).  In order to test this dependence, we show
also the results of the smallest cell length size $l$ of the highest
resolution grid interior to $D=1.0$.
The results of our numerical simulation clearly show a jump in the
value of $\ln \Lambda$ at $D=1.0$ when the cluster with a point-mass
potential enters the simulation area with the highest resolution.  We
applied the following fitting formula to the data
\begin{eqnarray}
  \label{eq:fitln1}
  \ln \Lambda & = &  \ln(b_{\rm max}) - \ln(b_{\rm min}) \nonumber \\
  & = & \ln (a \cdot D) + b
\end{eqnarray}
where $a$ and $b$ are two constants.  If $b_{\rm min} = b_{\rm
  min,theo} + l$ the values of $b$ would be $2.458$ in the region
$D>1.0$ and $3.829$ in the region $D<1.0$.  Using again the NLLS
algorithm of {\it gnuplot} we obtain the following results for $a$ and
$b$ (values for $D<1.0$ are given in brackets):
\begin{eqnarray}
  \label{eq:fitln2}
  a & = & 1.92 \pm 0.03 \ \ (2.68 \pm 0.06) \\
  \label{eq:fitln3}
  b & = & 2.46 \pm 0.02 \ \ (3.83 \pm 0.02).
\end{eqnarray}
The values for $b$ are in very good agreement with the predicted
values derived from Eq.~\ref{eq:bmin}.  The values for $a$ differ from
the results of $a'$ and the study by \citet{spi03}, because in these
cases a global $\ln \Lambda$ was used to fit the entire orbital decay
curve.  The two independent fitting curves are shown as the dashed
(dotted) line in Fig.~\ref{fig:i19-fit}.  We use the dashed fitting
line for the examination of the results in the live star-cluster
simulations.  The values for $\ln \Lambda$ derived for a cluster with
a point-mass potential with the given resolution of the code should be
systematically larger than the values of $\ln \Lambda$ expected for an
extended star cluster but at least we get an upper limit for $\ln
\Lambda$ which should provide us with a lower limit on the mass taking
part in the dynamical friction process.

\subsection{A compact star cluster with a size smaller than its tidal
  radius} 
\label{sec:i21}

\begin{figure}
  \begin{center}
    \epsfxsize=4.0cm \epsfysize=4.0cm \epsffile{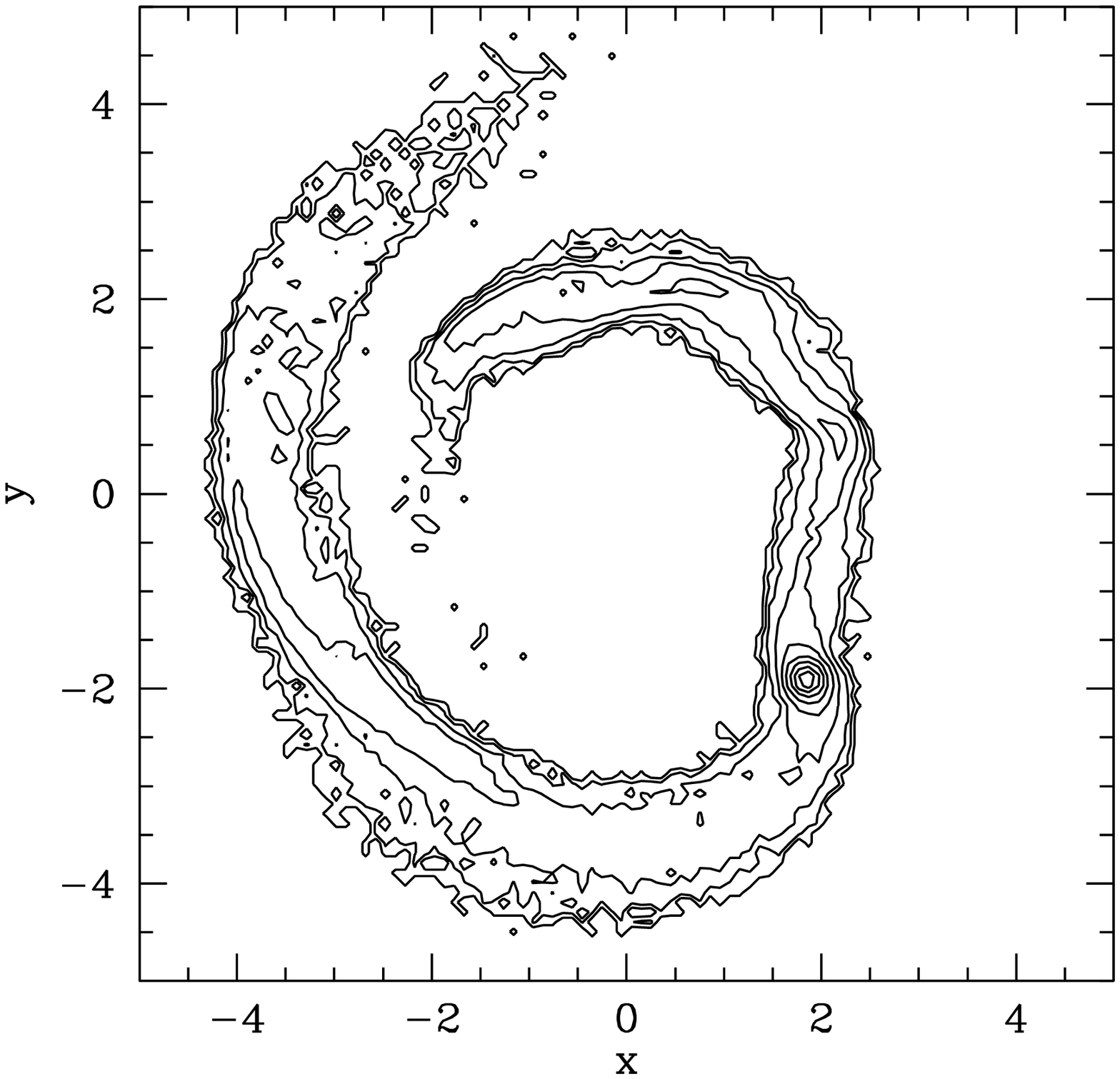}
    \epsfxsize=4.0cm \epsfysize=4.0cm \epsffile{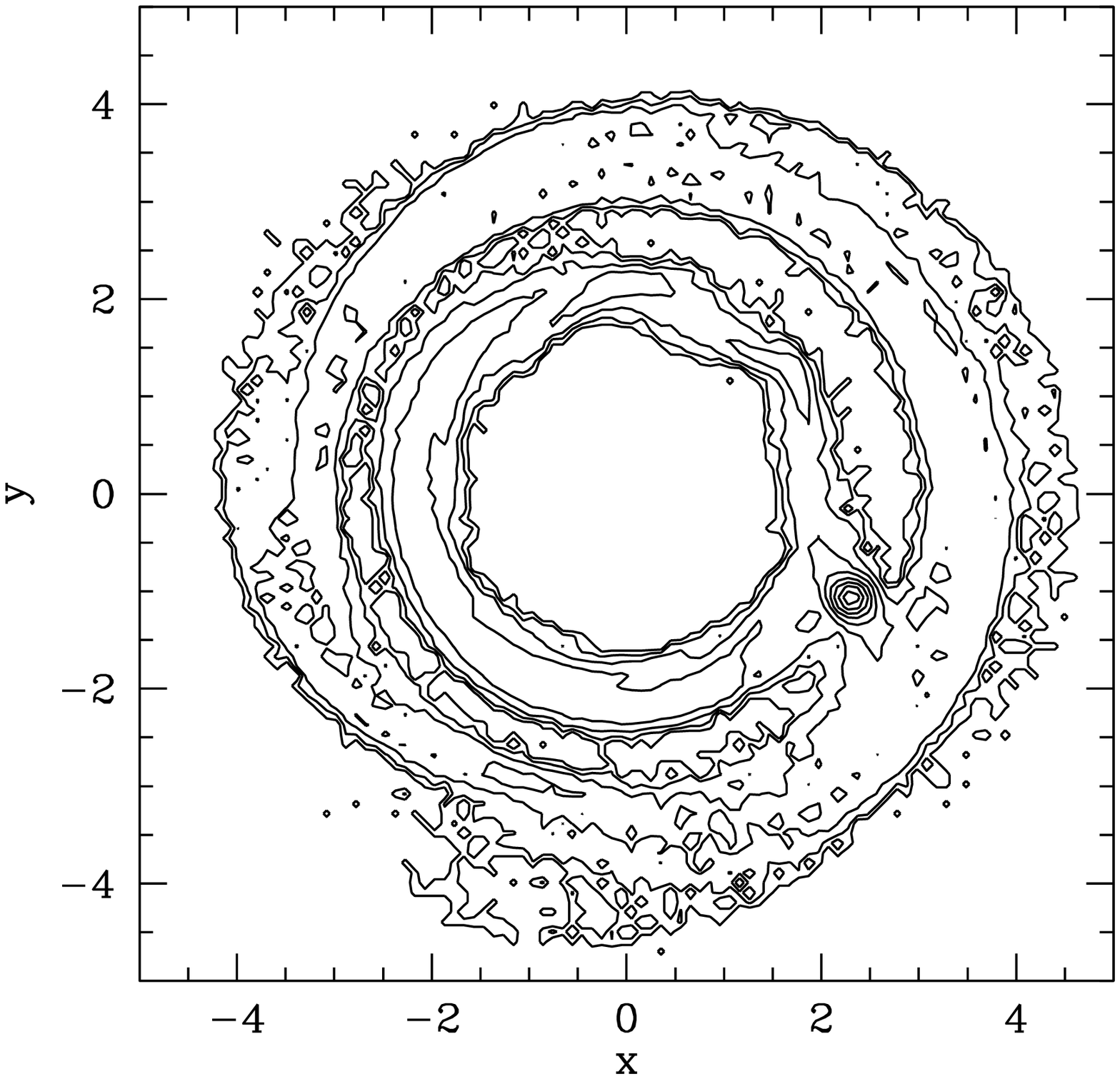}
    \epsfxsize=4.0cm \epsfysize=4.0cm \epsffile{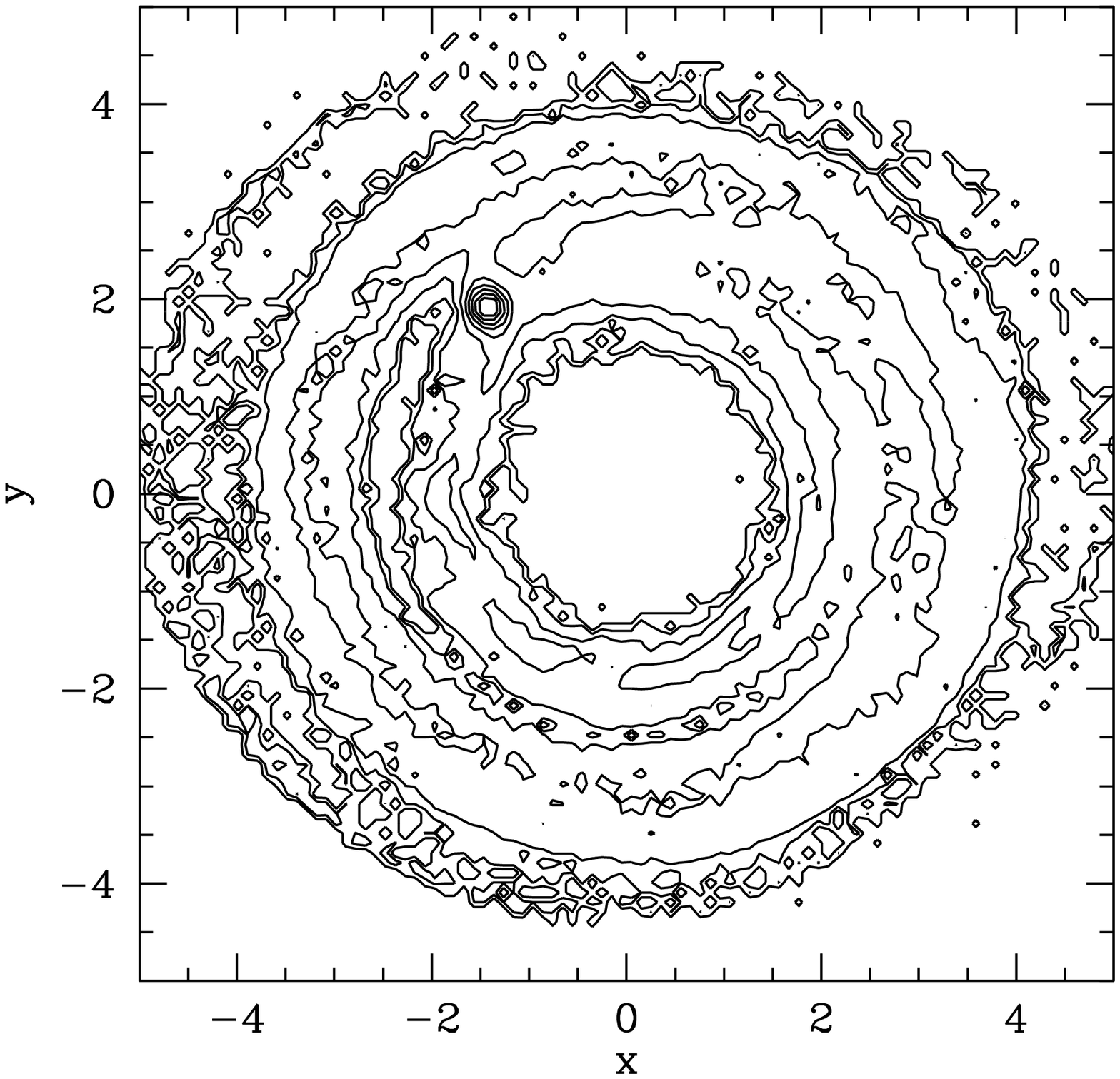}
    \epsfxsize=4.0cm \epsfysize=4.0cm \epsffile{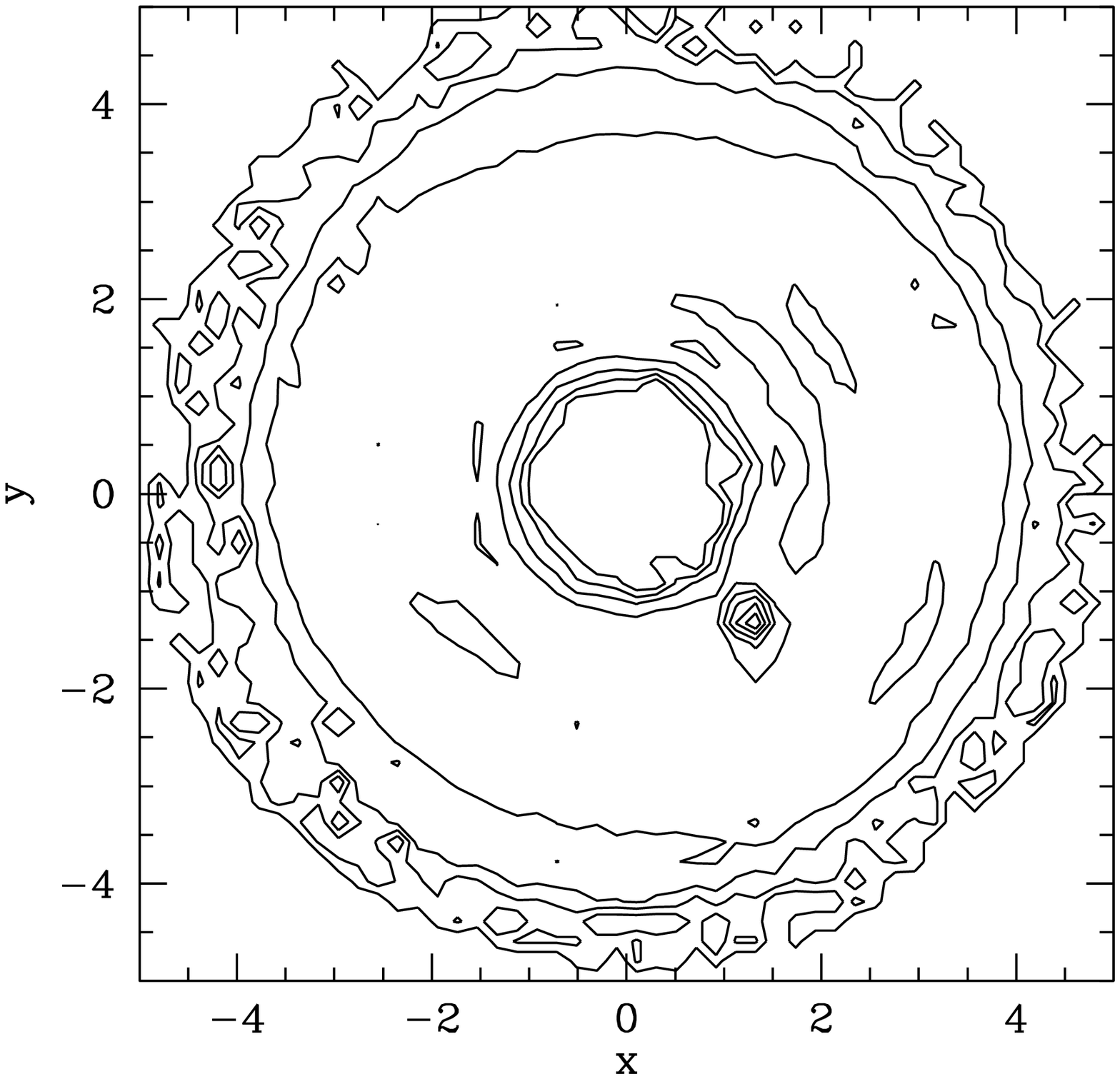}
    \epsfxsize=4.0cm \epsfysize=4.0cm \epsffile{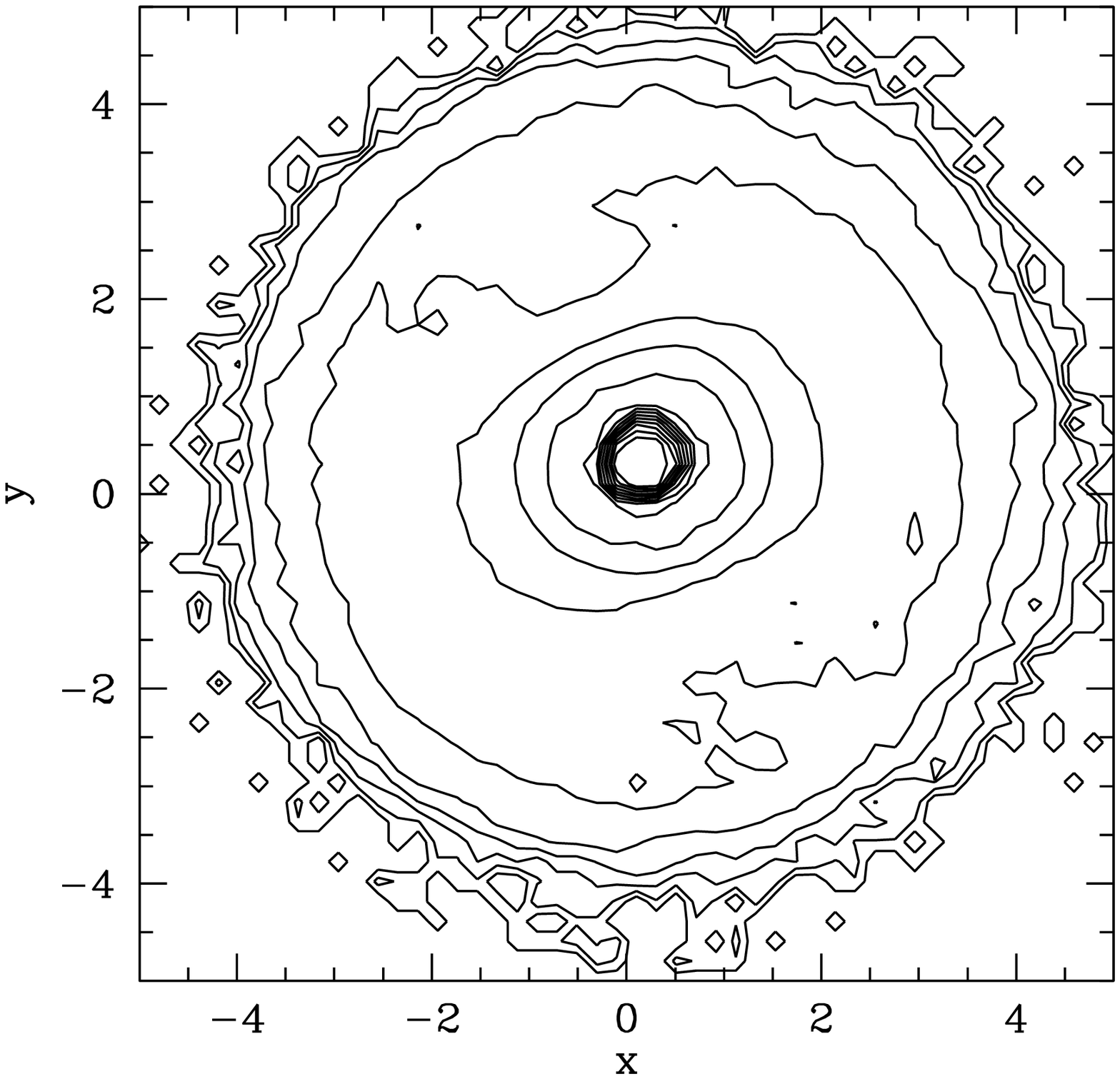}
    \caption{Contour-plots of the time-evolution of the sinking star
      cluster (smaller than the tidal radius).  Shown are the
      time-steps 2.5, 5.0, 10.0, 20.0 and 40.0.} 
    \label{fig:i21-cont}
  \end{center}
\end{figure}

\begin{figure}
  \begin{center}
    \epsfxsize=7.5cm 
    \epsfysize=7.5cm 
    \epsffile{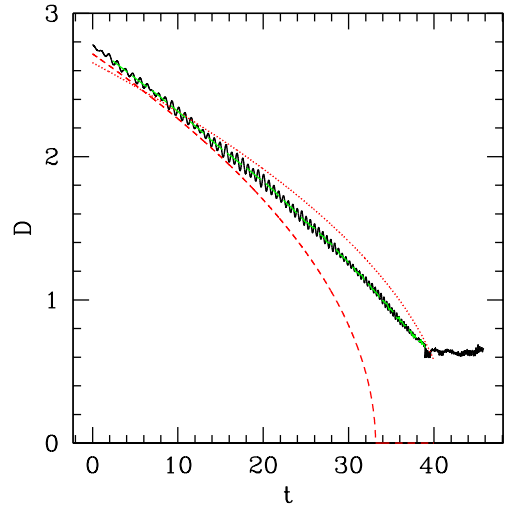}
%%%    \epsffile{i21-radfit.eps}
    \caption{The orbital decay of a compact stellar cluster which is
      smaller than its tidal radius.  Shown is the distance $D$ of the
      star cluster to the centre of the background vs.\ time (solid
      line; black on-line).  The short dashed line (red on-line) shows
      the fitting curve derived for the point-mass case ($M_{\rm cl} =
      1.0 = const.$) and the dotted line (red on-line) shows a fit of
      Eq.~\ref{eq:fitrad2} to the whole curve, i.e.\ trying to fit a
      constant $M_{\rm cl} \ln \Lambda$.  As expected these two curves
      do not fit the data properly.  The long dashed line (green
      on-line) which is barely visible because it fits the data almost
      exactly is again Eq.~\ref{eq:fitrad2} but this time the actual
      values for $M_{\rm cl} \ln \Lambda$, as determined below, are
      inserted.}  
    \label{fig:i21-rd}
  \end{center}
\end{figure}

\begin{figure}
  \begin{center}
    \epsfxsize=07.5cm 
    \epsfysize=07.5cm 
    \epsffile{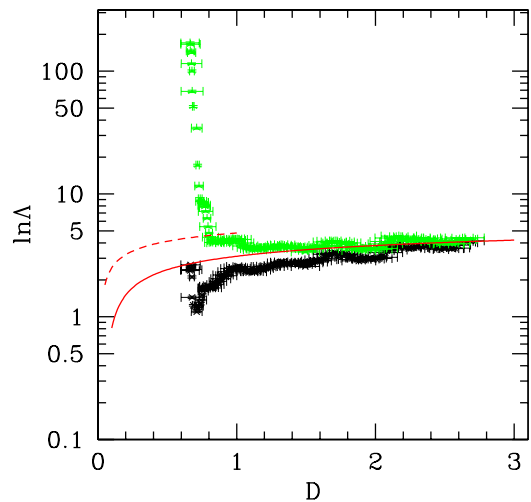}
%%%    \epsffile{i21-lambdafit.eps}
    \caption{The magnitude of $\ln \Lambda$ associated with the
      orbital decay of a compact stellar cluster due to dynamical
      friction.  The values of $\ln \Lambda$ are derived for each 5
      oscillations.  Crosses (black on-line) are deduced under the
      assumption of a constant total mass ($M_{\rm cl} = 1.0$) of the
      cluster (see Eq.~\ref{eq:l1}).  The tri-pods (green on-line) are
      derived under the assumption that only the bound mass contribute
      to the dynamical friction (see Eq.~\ref{eq:l2}).  The solid and
      dashed lines (red on-line) are the fitting curves for $\ln
      \Lambda$ derived from the point-mass potential case.}
    \label{fig:i21-fit}
  \end{center}
\end{figure}

\begin{figure}
  \begin{center}
    \epsfxsize=07.5cm 
    \epsfysize=07.5cm 
    \epsffile{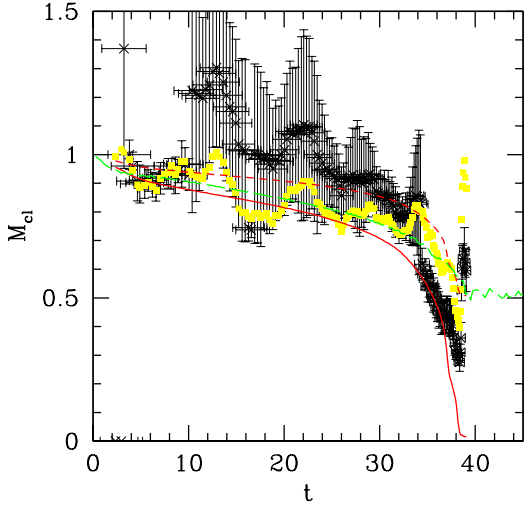}
%%%    \epsffile{i21-massfit.eps}
    \caption{Mass evolution of the star cluster smaller than the tidal
      radius.  Solid line (red on-line) is the bound mass of the
      cluster as calculated by {\sc Superbox}.  Crosses (black) and
      squares (yellow) denote the mass of the cluster which takes part
      in the dynamical friction if we insert the point-mass values of
      $\ln \Lambda$ in Eq.~\ref{eq:fitrad2} and solve for the cluster
      mass.  For the crosses we take the measured values; for the
      squares the values derived with the fitting curve from
      Fig.~\ref{fig:i19-fit}.  The short dashed line (red) shows the
      bound mass of the cluster plus the mass which is spread along
      the orbit in a torus of the size of the cluster around the
      galaxy and only particle with the same velocity signature
      (orbital velocity plus/minus initial velocity dispersion) as the
      cluster are taken into account.  Long dashed line (green) is our
      rule-by-thumb: bound mass plus half the unbound mass.} 
    \label{fig:i21-mass}
  \end{center}
\end{figure}

Typical satellites have extended mass distribution.  For example, the
internal density distribution within dwarf spheroidal galaxies is
relatively flat.  The half mass radii of many loosely bound globular
clusters are significant fraction of their tidal radii.  Globular
clusters near the Galactic centre such as the Arches cluster also have
relatively flat density profile.  The characteristic velocity
dispersion within these systems are closer to that of truncated
isothermal sphere than that for a point mass potential.  In order to
simulate a more realistic potential for typical satellites, we adopt a
model in which a cluster is represented by $10^6$ particles with a
Plummer-model phase-space distribution.

We place this star cluster at the same half-mass radius of the
background galaxy as in the previous simulation for a cluster with a
point-mass potential.  We deduce the magnitude of the cluster's tidal
radius analytically using formula 7-84 from Binney and Tremaine
(1987).  We set the Plummer radius to be $r_{\rm pl}= 0.05$. With a
total mass $1.0$, the analytic tidal radius at $D_{0}$ is $\sim 0.25$.
In this simulation, our objective is to examine whether the magnitude
of $\ln \Lambda$ for a realistic cluster potential differ
significantly from that for a point-mass potential.  In order to make
this comparison with the previous model, we consider a cluster which
is expected to be well preserved at least for several galactic orbital
periods.  In order to guarantee that the total mass of this cluster is
initially within its tidal radius, we introduce a cut off in the
stellar density at $r_{\rm cut} = 0.25$.

The results of this simulation at five epochs are shown in
Fig.~\ref{fig:i21-cont} (see also Fig.~\ref{fig:i21-mass}).  These
figures indicate that, due to tidal heating, the cluster looses mass
only slowly with time.  The cluster is therefore able to migrate
deeply into the potential of the host galaxy.  Fig.~\ref{fig:i21-rd}
shows the evolution of the star cluster distance from the centre, $D$.
The dotted line shows a fit to the orbit-decay curve for all
data-points, while the dashed line gives the fitting line which is
derived from the cluster model with a point-mass potential.  It is
evident that none of these lines provides an excellent approximation
of the data-points. 

Following the same procedure as in the model with the point-mass
potential, we now determine the combined values of $M_{\rm cl} \ln
\Lambda$ for the small time-slices.  While the bound mass, the time
and distance of the cluster are instantaneous quantities which are
determined at each time-step, the combined quantity $(M_{\rm cl} \ln
\Lambda) (t)$, determined via the sinking rate using
Eq.~\ref{eq:fitrad2}, can only be accessed in a time-average
approximation.  Although all quantities evolve over finite time-spans,
we evaluate the magnitude of $(M_{\rm cl} \ln \Lambda) (t)$ with the
mean values for the time, distance and bound mass.  An optimised time
interval is chosen so that it is sufficiently short to ensure that the
instantaneous quantities do not evolve significantly while it is
adequate to limit the uncertainties in the estimated sinking rate
during each of these time intervals.  This simple-to-use approach
introduces modest error bars in several figures throughout our
manuscript.

The results are plotted in Fig.~\ref{fig:i21-fit}.  The crosses
represent the value of $\ln \Lambda$ obtained under the assumption of
constant $M_{\rm cl} = 1.0$: 
\begin{eqnarray}
  \label{eq:l1}
  \ln \Lambda (t)_{\rm crosses} & = & (M_{\rm cl} \ln \Lambda) (t) \ /
  \ M_{\rm bound} (t=0)  
\end{eqnarray}
In Fig.~\ref{fig:i21-fit}, the crosses fall slightly below the fitting
line for the cluster with a point mass potential (which is represented
by the dashed line).  This disparity is expected since an extended
object should have a larger $b_{\rm min}$ than that with a point-mass
potential.  For $t<30$ or $D>1$, the difference between these two
simulation is less than 20\%.  But, it is also visible that the
deviation from the fitting line grows with time, especially at $t>30$
(or equivalently as $D$ decreases below 1).  This growing difference
is due to lost of mass from the stellar cluster.  This divergence
shows that a constant $M_{\rm cl}$ approximation does not adequately
represent the results of the simulation.

The tri-pods in the same figure represent the values of $\ln \Lambda$
derived under the assumption that only the bound mass is responsible
for dynamical friction:
\begin{eqnarray}
  \label{eq:l2}
  \ln \Lambda (t)_{\rm tri-pods} & = & (M_{\rm cl} \ln \Lambda) (t) \
  / \ M_{\rm bound} (t)  
\end{eqnarray}
The bound mass of the cluster is a quantity which is computed by {\sc
  Superbox}, the simulation programme used in this study, at each
time-step of the simulation.  In this routine an energy argument
is used instead of determining the Roche radius.  All particles having
negative energy with respect to the cluster potential are flagged as
bound.  Measuring $\ln \Lambda$ according to Eq.~\ref{eq:l2} gives 
values which are systematically above the fitting line representing
the cluster with a point-mass potential.  But, using the above
argument that an extended object should have a larger $b_{\rm min}$
than that with a point-mass potential, the tri-pods measured from this
simulation would be systematically below the fitting line if the bound
stars adequately account for all the mass which contributes to the
dynamical friction.  This disparity is a first hint that more
particles may take part in the dynamical friction than just the bound 
stars.  In the later stages of the evolution these values for $\ln
\Lambda$ increase quite dramatically which is a clear sign that
$M_{\rm cl}$ is underestimated.

In Fig.~\ref{fig:i21-mass}, we plot the bound mass of the object as a
function of time (solid line).  In the same figure, crosses and
squares represent the mass of the cluster taking part in the dynamical
friction process if we assume the same $\ln \Lambda$ as derived
for a cluster with a point-mass potential and solve for $M_{\rm cl}$
with Eq.~\ref{eq:fitrad2}:
\begin{eqnarray}
  \label{eq:mfit}
  M_{\rm cl}(t) & = & (M_{\rm cl} \ln \Lambda) (t) \ / \ (\ln
  \Lambda)_{\rm point mass} (t)
\end{eqnarray}
For the crosses we apply the actual values from the point-mass
simulation while the data-points of the squares are derived using the
smoothed fitting curve for $\ln \Lambda(D)$ from Eq.~\ref{eq:fitln1}.
(Since we have already shown that the magnitude of $\ln \Lambda(D)$
for a cluster with a Plummer potential is smaller than that for a
point mass potential, the actual total mass which contributes to the
dynamical friction is slightly larger than both the values represented
by the crosses and the squares.).  Even though the uncertainties are
large the data points show that the total mass which
contributes to the effect of dynamical friction is systematically above
the bound mass in the bound mass curve. 

In addition to the bound mass we determine the lost mass of the
cluster which is located in a ring of the cluster's extension around
the galaxy at the same distance and only the particles with the same
velocity signature as the cluster (orbital velocity $\pm$ velocity
dispersion) are counted.  Adding this mass to the bound mass is shown
as the short dashed line in Fig.~\ref{fig:i21-mass}.  This mass
estimate seems to fit the data much better than the bound mass only.
In an impulse approximation, the escaped cluster particles on the
opposite of the galaxy do not contribute to the dynamical friction and
should not be included in the active mass associated with the cluster.
But, the cluster does have follow up encounters with the perturbed
particles during either the present or subsequent orbits.  The
numerical results presented here is indicative that dynamical friction
is a collective effect and not only the bound mass is responsible for
it. To what extend the particles with large azimuthal separations from
the cluster contribute to its orbital decay needs to be addressed in
much more elaborate studies which may include alternative formulations
(other than impulse approximation) of dynamical friction.  Such an
analysis may provide a weighting function to appropriately include the
relatively weak contribution of the more distant particles.

We note a large fraction of the co-moving unbound mass is contributed
by particles which became detached from the cluster in the previous
orbit and are currently located within 2-3 Roche radii from the
cluster.  But attempts to use radial criteria like all particles
within 2, 3 or 5 tidal radii or counting only leading/trailing arm
particles (curves are not shown) were not successful in fitting the
data.  Thus, we applied a simple rule-by-thumb by adding half of the
unbound mass to the bound mass (shown as long dashed line in
Fig.~\ref{fig:i21-mass}).  This simple estimate fits the data nicely
taking into account that the 'actual' $\ln \Lambda$ of an extended
object should be smaller than the one of a point mass, i.e.\ the data
points have to be regarded as lower limits.  But please keep in mind
that this expression is just an {\it ad hoc} estimate without any
rigorous derivation.

\subsection{Diffuse star cluster with its cut-off radius larger than its 
tidal radius}
\label{sec:i20}

\begin{figure}
  \begin{center}
    \epsfxsize=4.0cm \epsfysize=4.0cm \epsffile{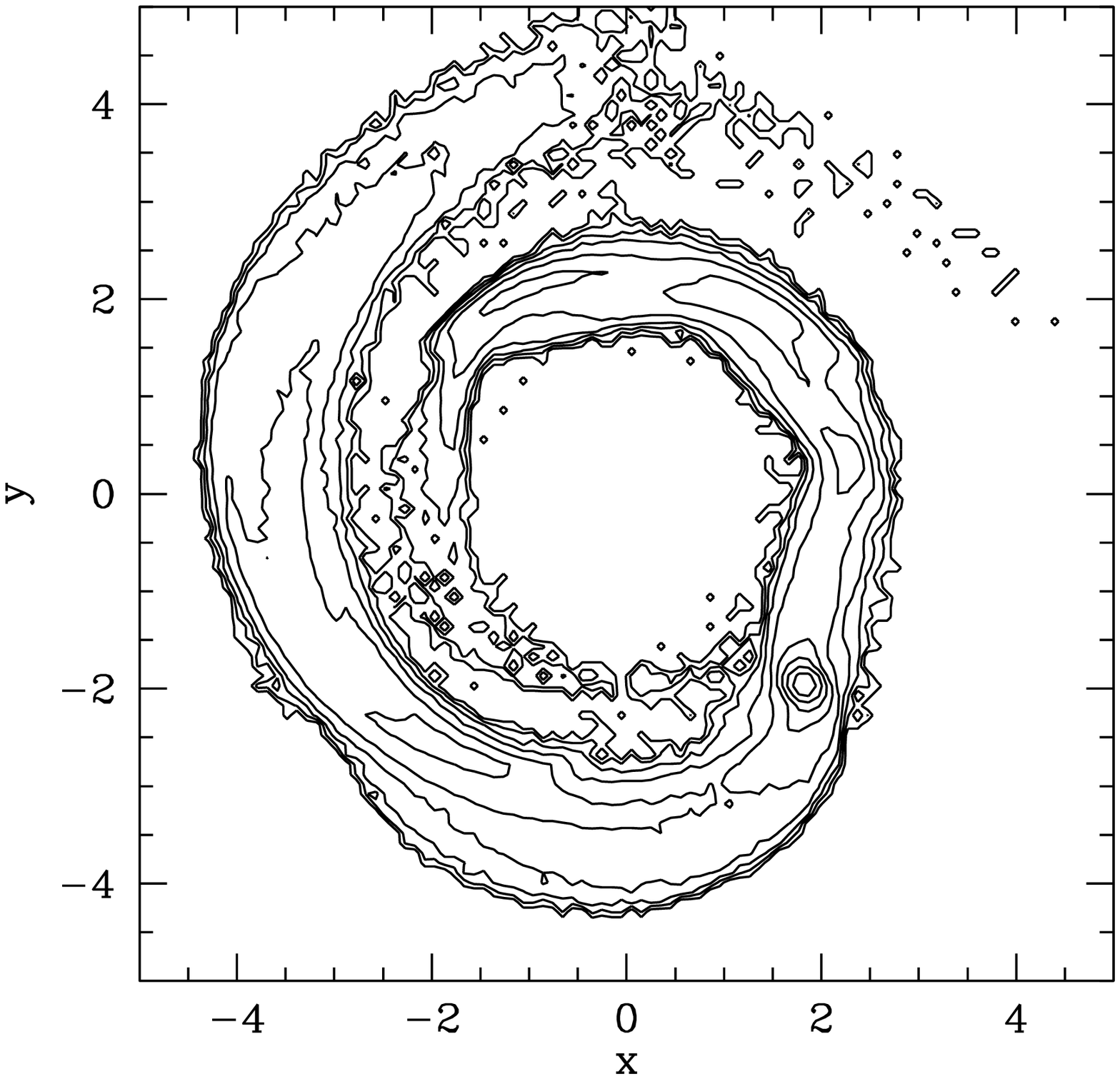}
    \epsfxsize=4.0cm \epsfysize=4.0cm \epsffile{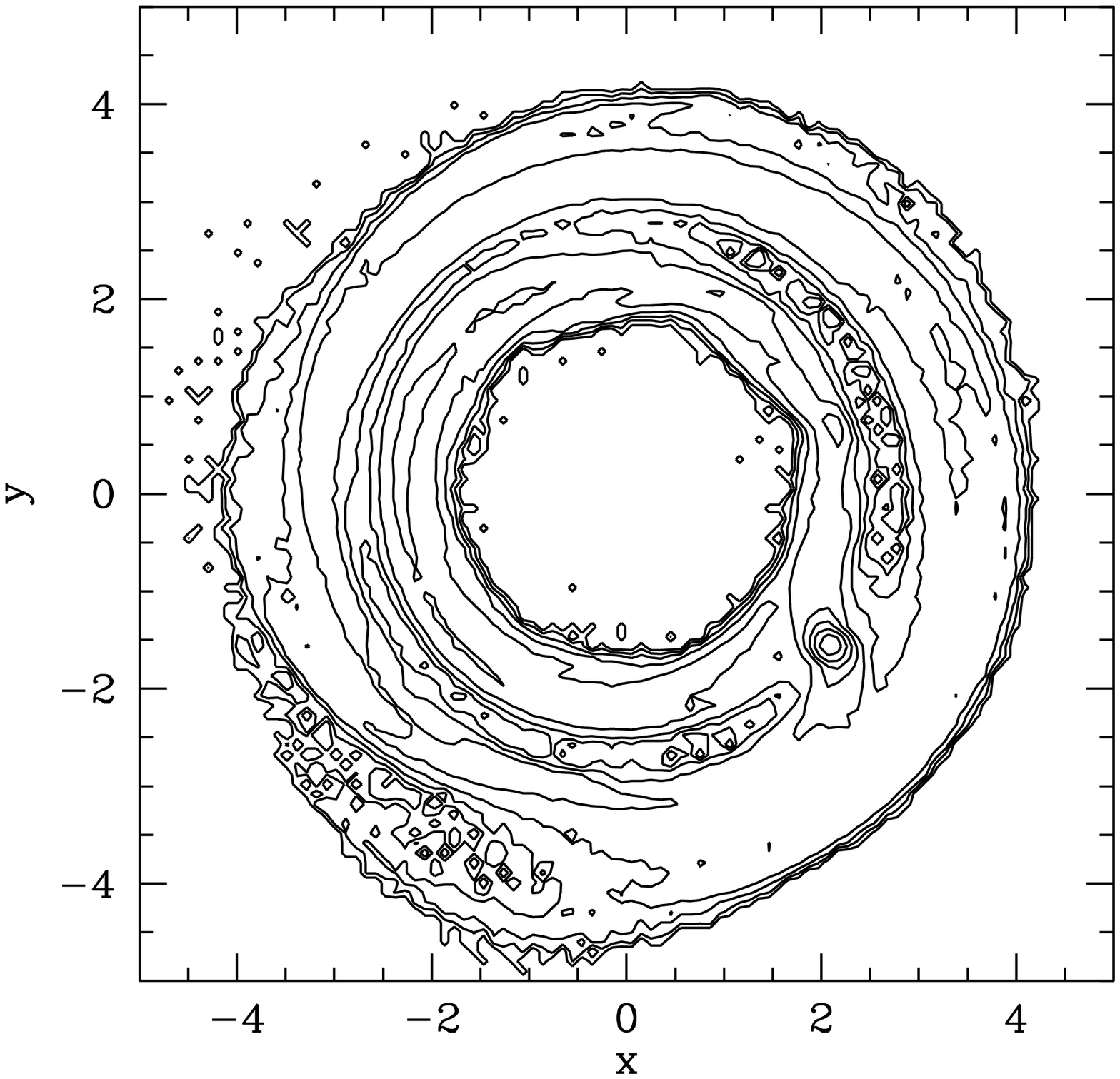}
    \epsfxsize=4.0cm \epsfysize=4.0cm \epsffile{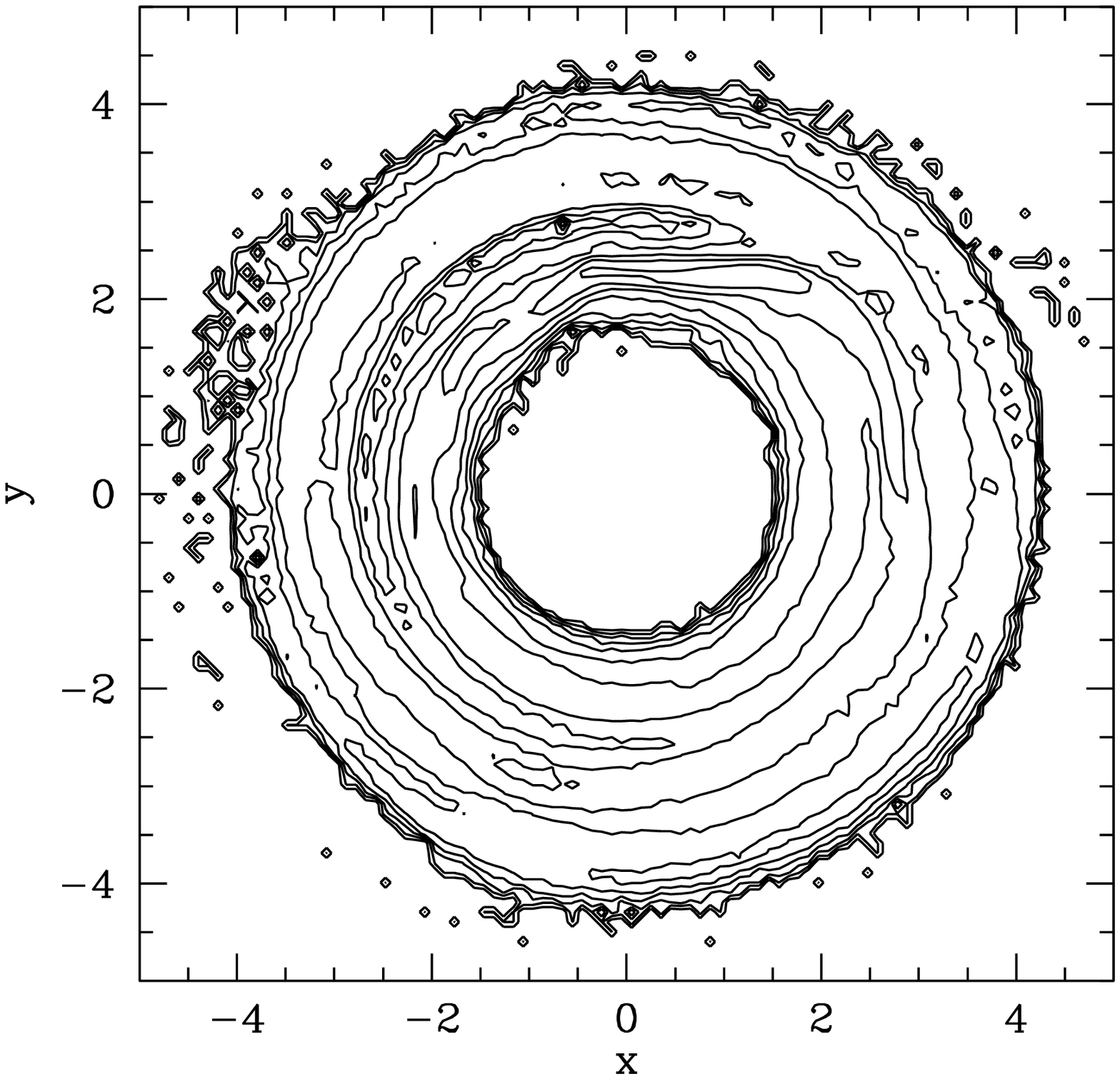}
    \epsfxsize=4.0cm \epsfysize=4.0cm \epsffile{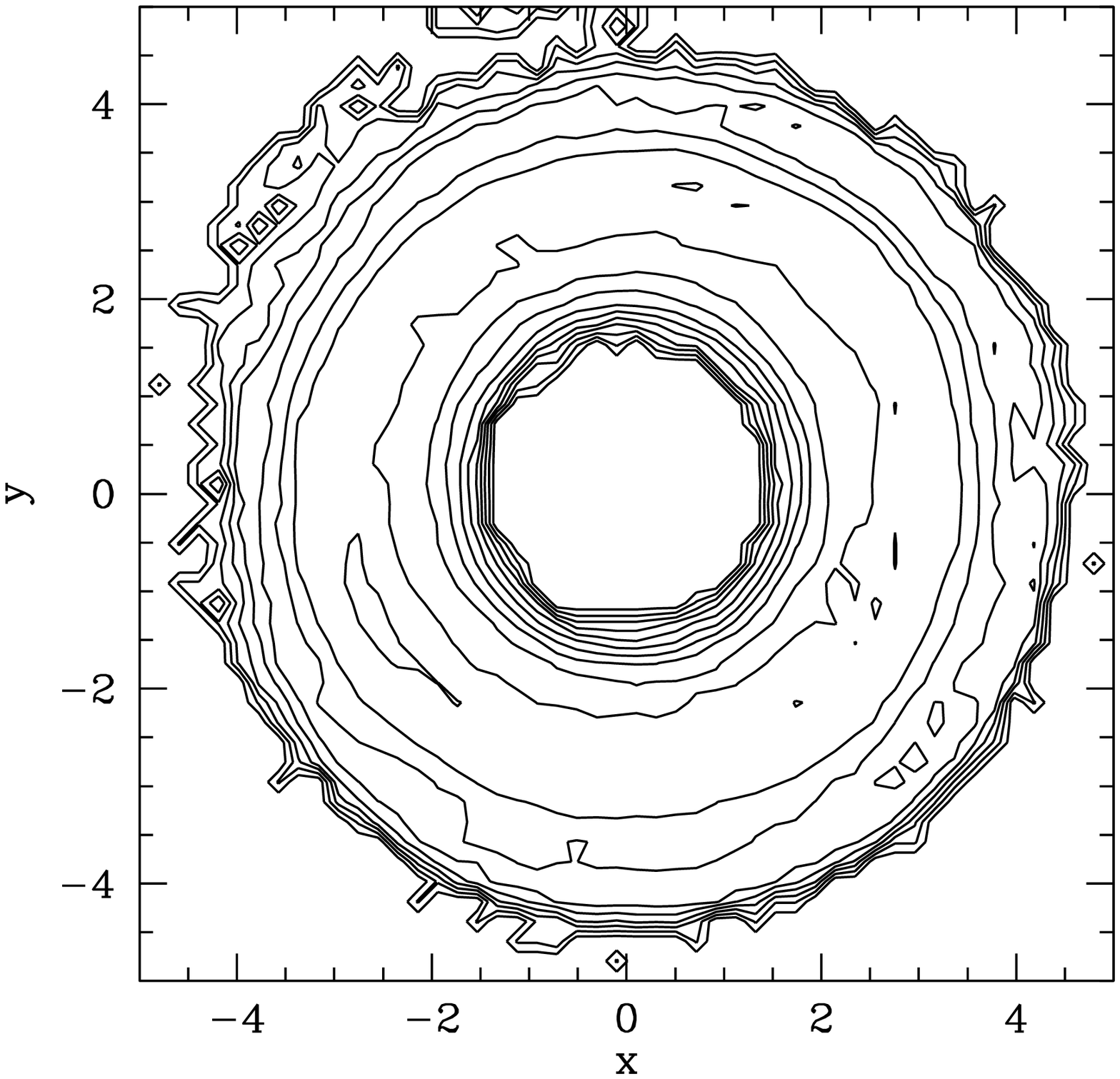}
    \caption{Contour plots of the time evolution of a star cluster
      larger than the tidal radius. Shown are the time-slices at 2.5,
      5.0, 10.0, and 20.0 time-units.}  
    \label{fig:i20-cont}
  \end{center}
\end{figure}

\begin{figure}
  \begin{center}
    \epsfxsize=07.5cm 
    \epsfysize=07.5cm 
    \epsffile{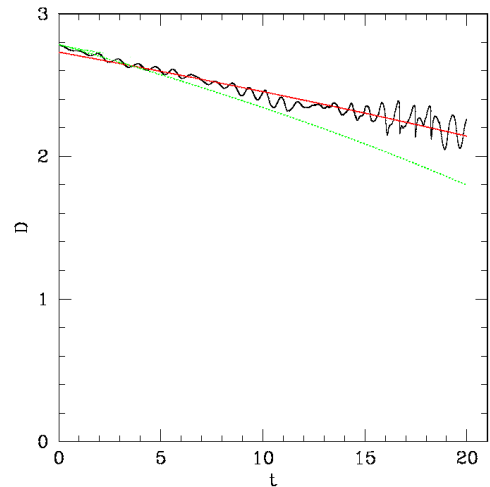}
%%%    \epsffile{i20-rd-new.eps}
    \caption{A diffuse star cluster which is larger than the tidal
      radius.  The orbital evolution of the star cluster is plotted
      as a solid line (black).  The dotted line (green) represents the
      maximum orbital decay rate derived for a cluster with a
      point-mass potential.  (Red) solid line is the orbital decay
      with the actual values for $M_{\rm cl} \ln \Lambda$ inserted.}
    \label{fig:i20-rd}
  \end{center}
\end{figure}

\begin{figure}
  \begin{center}
    \epsfxsize=07.5cm 
    \epsfysize=07.5cm 
    \epsffile{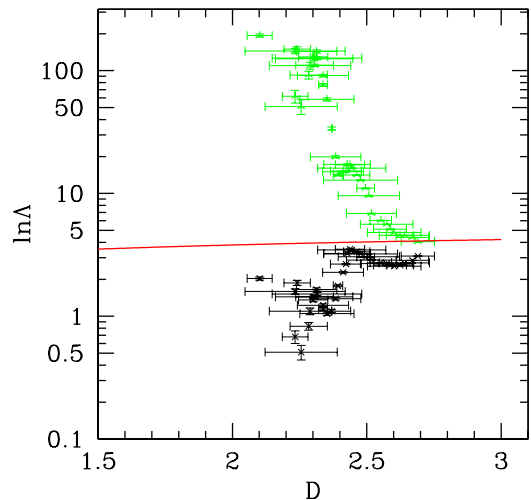}
%%%    \epsffile{i20-lambdafit.eps}
    \caption{The deduced values of $\ln \Lambda$ from the simulation
      of a diffuse cluster with a cut-off radius larger than its tidal
      radius.  Crosses (black) represent the values of $\ln \Lambda$  
      according to Eq.~\ref{eq:l1} and tri-pods (green) represent the
      values for $\ln \Lambda$ deduced from Eq.~\ref{eq:fitrad} under
      the assumption that $M_{\rm cl} = M_{\rm bound}$
      (Eq.~\ref{eq:l2}).  The fiducial values of $\ln \Lambda$ for a  
      cluster with a point-mass potential is also plotted for
      comparison purpose (red solid line).  These data show that the 
      assumption the bound mass clearly underestimates the total mass
      responsible for dynamical friction because the values of $\ln
      \Lambda$ increase to unphysical values.}
    \label{fig:i20-fit}
  \end{center}
\end{figure}

\begin{figure}
  \begin{center}
    \epsfxsize=07.5cm 
    \epsfysize=07.5cm 
    \epsffile{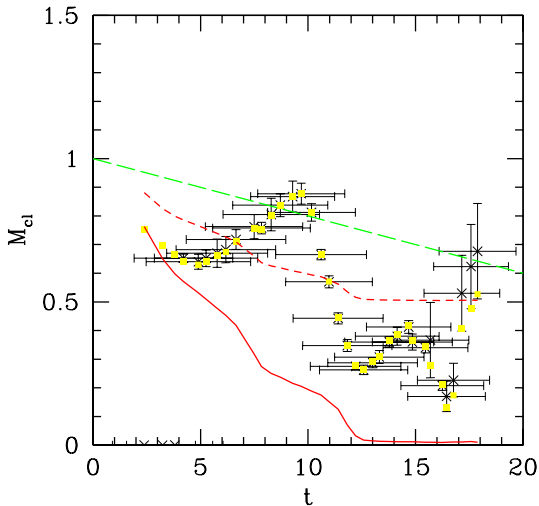}
%%%    \epsffile{i20-massfit.eps}
    \caption{Mass evolution of a star cluster with an initial cut-off
      radius larger than its tidal radius.  Solid line (red) is the
      bound mass as calculated by {\sc Superbox}.  Again the crosses
      (black) and squares (yellow) are the values for the mass taking
      part in the dynamical friction process if $\ln \Lambda$ deduced
      for a cluster with a point-mass potential is applied (for
      explanation see Fig.~\ref{fig:i21-mass}).  Long dashed line
      (green) represents all the stars in a torus with a width
      comparable to the initial size of the cluster at the distance of
      the cluster which have the same velocity signature as the stars
      within the cluster.  Short dashed line (red) corresponds to the
      value of effective mass in accordance with our rule-by-thumb.}
    \label{fig:i20-mass}
  \end{center}
\end{figure}

Toward the end of the previous simulation, the compact cluster migrate
toward the centre of the background galaxy where the external tidal
perturbation is sufficiently strong to cause total disruption.  In
order to study the effect of the dynamical friction during the
disruption of the cluster, we carry out a third simulation with a star
cluster having a stellar distribution extending beyond it tidal
radius.  For the initial cluster structure, we adopt a Plummer-sphere
with Plummer-radius of $r_{\rm pl} = 0.1$ and a cut-off radius of
$r_{\rm cut} = 0.5$.  This cluster is launched on the same circular
orbit as the previous simulations.  It loses mass immediately and is
destroyed after a few orbits around the background galaxy.  With this
set of initial conditions, we exaggerate the rate of tidal disruption
of the cluster.  Although this set of initial conditions seems to be
unattainable, in some clusters (such as the Arches cluster), where the
internal two-body relaxation may sufficiently intense, the outer
regions of the clusters may be replenished on time scale shorter than
their galactic orbital periods \citep{mac02}.  Most dwarf galaxies are
on highly eccentric orbits with period comparable to their internal
evolution time-scales.  These systems tend to lose bound mass near
their perigalactica where the tidal radius may also be reduced inside
their cut-off radius.  If this perigalacticon passage is brief
compared to the internal time-scales the object does not have time to
adjust smoothly to the mass-loss and is therefore in a similar state
as our model.

In Fig.~\ref{fig:i20-cont} we illustrate the orbital evolution of the
star cluster.  This figure clearly shows that already at time 10.0
there is no longer any distinctive density enhancement at the star
cluster position.  Nevertheless, we are able to determine the centre
of density of the last percents of bound particles and trace its orbit.
Fig.~\ref{fig:i20-rd} shows that the remains of the object still sink
to the centre even after the tidal disruption of the cluster is nearly
completed.  Beyond time 15.0 the determination of the centre of
density of the residual star cluster becomes very difficult and highly
uncertain.  But there still are indications that the remnant core
continues to undergo orbital decay even after its near complete
disruption.

In Fig.~\ref{fig:i20-fit} it is visible that in this case the
assumption that only the bound mass is responsible for dynamical
friction is no longer valid.  Instead of staying close to the values
determined in the point-mass case the values of $\ln \Lambda$
according to Eq.~\ref{eq:l2} keep on increasing to really unphysical
values of more than 200.

Following the analysis for the previous model, we use the value of
$\ln \Lambda$ which was determined for a cluster with a point-mass
potential to derive a mass which is responsible for the dynamical
friction.  This mass-estimate is shown in Fig.~\ref{fig:i20-mass} as
crosses and squares.  The values of this mass are clearly above the
line of the bound mass and below the line of the particles inside the
ring of the star-cluster orbit with the same velocity-signature.  They
approximately correspond to the bound mass plus half of the mass which
is already lost (our rule-by-thumb).

\subsection{A moving cluster without self-gravity}
\label{sec:nsg}

\begin{figure}
  \centering
  \epsfxsize=07.5cm 
  \epsfysize=07.5cm 
  \epsffile{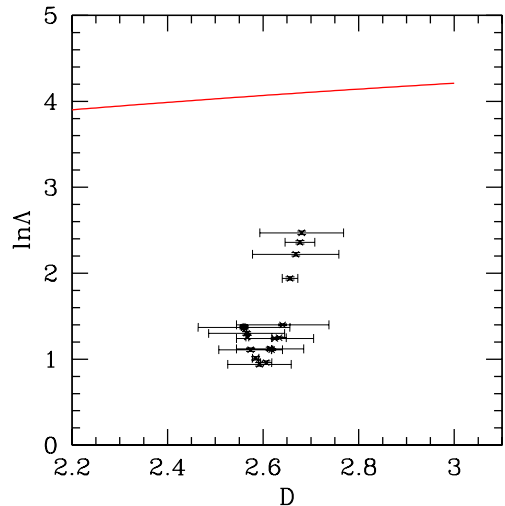}
%%%  \epsffile{nfc-lambdafit.eps}
  \caption{A moving cluster of unbound stars which only participate in
    gravitational interaction with the background stars.  The mutual
    gravity between the cluster stars is neglected. (Therefore the
    cluster stars have no internal velocity dispersion but follow only
    the orbital motion at the beginning of the simulation).  This
    figure shows the fitted values for $\ln \Lambda$ according to
    Eq.~\ref{eq:l1} for each 5 oscillations as function of the
    distance $D$ to the centre of the background galaxy.  Only a few
    data points are measurable before the cluster disperses
    completely.  The data-points are well below that for a cluster
    with point-mass potential, which is shown as solid line (red).
    But the magnitude of $M_{\rm cl}\ln\Lambda$ is definitely not
    zero.}  
  \label{fig:nfcirc}
\end{figure}

\begin{figure}
  \centering
  \epsfxsize=07.5cm 
  \epsfysize=07.5cm 
  \epsffile{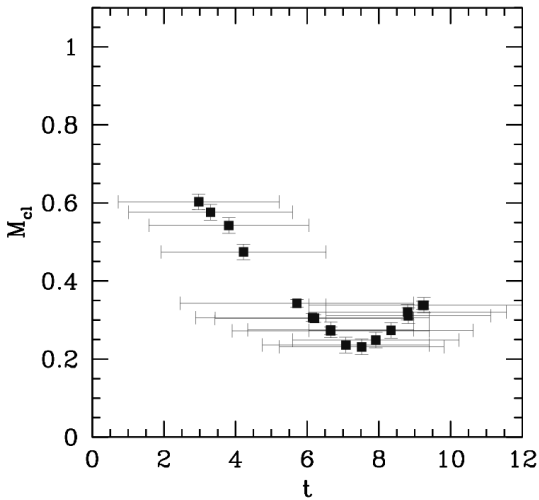}
%%%  \epsffile{nfc-massfit.eps}
  \caption{Applying $\ln \Lambda$ from the point mass simulation and
    calculating the mass taking part in the dynamical friction
    process as a function of time.  In the beginning the data points
    are spread around half of the cluster mass, which is a
    verification of our rule-by-thumb (Here the whole mass of the
    cluster is unbound from the onset of the simulation, therefore the
    mass for dynamical friction should be half of the cluster mass.).
    At later times the moving cluster disperses more and more and
    the values start to drop.}
  \label{fig:nfc2}
\end{figure}

We provide additional evidence that the recently-detached stars
contribute to the process of dynamical friction.  We construct an
artificial demonstration in which the mutual gravity among the stars
associated with the cluster is neglected.  In this construct, none of
the cluster particles are bound to each other but the gravitational
interaction between them and the background stars is preserved.

With this prescription, the cluster stars begin to disperse almost
immediately.  But, we are able to follow the orbital evolution of the
density centre for several orbits.  Following the previous analysis,
we determine $M_{\rm cl} \ln \Lambda$ and plot the derived values
against the mean distance to the centre in Fig.~\ref{fig:nfcirc}.
These values are well below the analogous values for the cluster with
a point-mass potential.  Based on the assumption that the simulation
for the cluster with a point mass potential provides a reliable value
of $\ln \Lambda$, we determine the mass $M_{\rm cl}$ responsible for
dynamical friction of this moving group in Fig.~\ref{fig:nfc2}.  Even
though the bound mass of this moving group is actually zero from the
onset of the simulation, the coherent movement of the cluster
particles can induce dynamical friction.  At the beginning
approximately half of the cluster mass is involved which proves again
our rule-by-thumb and only later the particles become too disperse and
this mass is reduced.

\subsection{Cluster on an Eccentric Orbit}
\label{sec:ecc}

\begin{figure}
  \centering
   \epsfxsize=07.5cm 
   \epsfysize=07.5cm 
   \epsffile{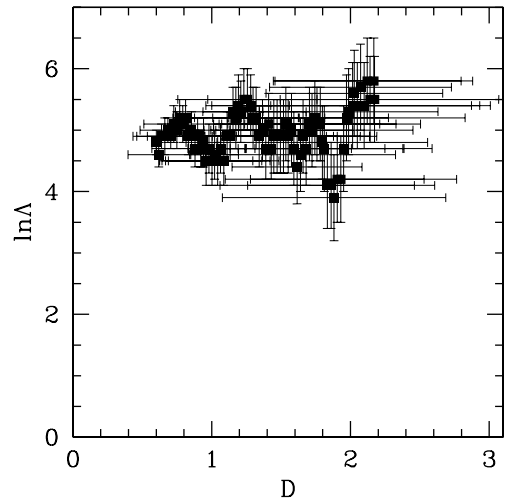}
%%%   \epsffile{pmex-lambdafit.eps}
    \caption{Point-mass star cluster on an eccentric orbit.  Shown are 
      the derived values for $\ln\Lambda$ in the point-mass case
      plotted against the mean distance of the cluster to centre of 
      the galaxy (according to Eq.~\ref{eq:decc}).  Because of the
      eccentricity of the orbit the uncertainties in $D$ are quite
      large (error-bars denote the maximum and minimum distance to
      centre during 5 oscillations which are used to determine
      $\ln \Lambda$).}
  \label{fig:eccentric}
\end{figure}

\begin{figure}
  \centering
   \epsfxsize=07.5cm 
   \epsfysize=07.5cm 
   \epsffile{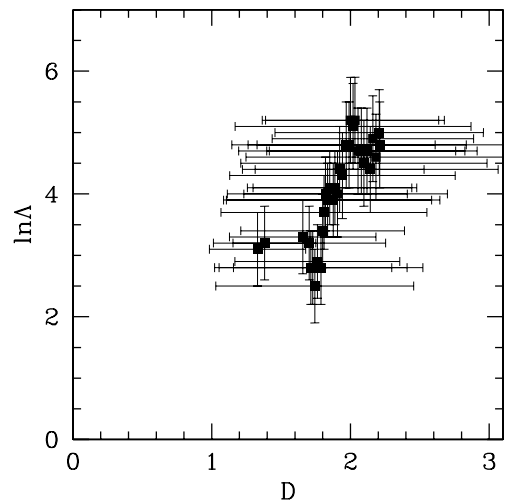}
%%%   \epsffile{clex-lambdafit.eps}
  \caption{Compact star cluster on an eccentric orbit.  Shown are the
    values for $\ln \Lambda$ for the star cluster.  The cluster used in
    this simulation is the identical model which is used for the
    cluster smaller than the tidal radius case with circular orbit.
    The values for $\ln \Lambda$ are derived if we assume $M_{\rm
      cl} =$ const.$= 1.0$. (see Eq.~\ref{eq:l1})}
  \label{fig:ecc1}
\end{figure}

\begin{figure}
  \centering
  \epsfxsize=07.5cm 
  \epsfysize=07.5cm 
  \epsffile{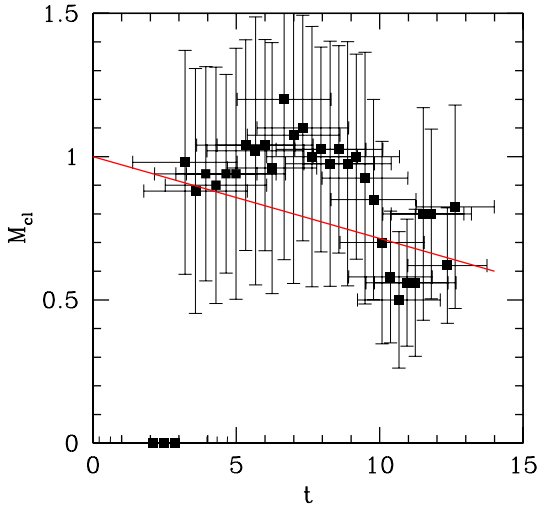}
  %%% \epsffile{clex-massfit.eps}
  \caption{Compact star cluster on an eccentric orbit.  Shown is the
    mass responsible for the dynamical friction if we adopt the values
    for $\ln \Lambda$ from the point mass case (eccentric orbit; see
    Fig.~\ref{fig:eccentric}).  Solid line (red) denotes the bound mass
    of the cluster.  The simulations with eccentric orbits were
    performed to investigate if the mass responsible for the dynamical
    friction shows similar 'oscillations' than in the circular orbit
    simulations.  The values still show the same kind of long-period
    oscillations, but they are now smaller than the actual
    uncertainties of the mass.} 
  \label{fig:ecc2}
\end{figure}

Finally we examine the nature of the oscillations which are apparent
in Figs.~\ref{fig:i21-mass}, \ref{fig:i20-mass} \&
Fig.~\ref{fig:nfc2}.  These oscillations do not show any preferred
period and are longer than and not related to the orbital time-scale
of the cluster.  These oscillations appear to be caused by the
enhanced dynamical friction associated with the close encounters
between the residual cluster and the stars which became detached
during previous orbits and have wrapped around the galaxy to the
vicinity of the cluster.  In our model setup this wrap-around
happens already in the very early stages (after 3-4 orbits) of the
simulations as shown in Figs.~\ref{fig:i21-cont} \&
\ref{fig:i20-cont}.  The trailing and leading debris have local
density variations which can lead to variations in their perturbation
on the motion of the residual clusters.  As a test for this
conjecture, we placed a cluster on an initial eccentric orbit.  In the
gravitational field of the background galaxy, the galactic orbit of
the cluster and its debris diverge due to differential precession and
the relatively large velocity differences between the residual cluster
and the wrapped around debris stream limits the influence of the
detached stars on the orbital evolution of the residual cluster. 

Following previous procedures, we first carried out a simulation with
a cluster with a constant point-mass potential.  The cluster orbit has
an initial eccentricity of $0.4$ and an orbital period of $0.74$.  We
derive the values of $\ln \Lambda$ as a function of cluster's mean
distance from the galactic centre, which is now computed as
\begin{eqnarray}
  \label{eq:decc}
  D & = & (\langle D_{\rm apo} \rangle + \langle D_{\rm peri} \rangle)
  / 2  
\end{eqnarray}
for each time-slice which covers 5 orbits around the centre of the
background.  A comparison between Fig.~\ref{fig:eccentric} and
Fig.~\ref{fig:i19-fit} indicates that the value of $\ln \Lambda$ for
the same value of $D$ is somewhat larger for a cluster with an
eccentric orbit than that with a circular orbit.  The modest epicycle
amplitude expands the radial range of halo stars which respond to the
gravity of the cluster with an eccentric orbit than that with a
circular orbit.  This effect corresponds to an increase in the
magnitude of $b_{\rm max}$ in Eq.~\ref{eq:lambda}.

Fig.~\ref{fig:ecc1} displays the corresponding values for a compact
star cluster with the same eccentric initial galactic orbit as that
for the point-mass potential test model.  We used the same Plummer
stellar phase space distribution for the compact cluster as described in
Sect.~\ref{sec:i21}.  In this case, the cluster's cut-off radius is
smaller or larger than its tidal radius near its initial apogalacticon
or perigalacticon respectively.  Under the assumption that all of
its initial mass contributes to the effect of dynamical friction, the
magnitude of $\ln \Lambda$ for the active cluster is smaller than that
of a cluster with a point-mass potential.  This disparity is
consistent with the results obtained in Sect.~\ref{sec:i21} and it is
particularly notable at relatively small $D$ where a significant
fraction of the cluster is lost.

Using the values of $\ln \Lambda$ derived for the cluster with a
constant point mass potential (see Fig.~\ref{fig:eccentric}), we
estimate the total cluster mass 
which contributes to the process of dynamical friction.  The results
are plotted as dots with error bars in Fig.~\ref{fig:ecc2}.  For
comparison, we also plotted evolution of the mass which remains bound
to the cluster.  This quantity is computed self consistently from the
relative total energy of each particle with respect to the cluster.
It is plotted as a solid (red on-line) line.  The bound mass appears
to be less than the total mass which contributes to the effect of
dynamical friction.  It is also notable that the amplitude of
long-term oscillation in $M_{\rm cl}$ is significantly
reduced.  Toward the end of the simulation (at $t>10$), the effective
mass also becomes comparable to the bound mass which may also be due
to the orbital divergence of the debris from the parent cluster.

\section{Conclusions}
\label{sec:conclus}

Dynamical friction is an important process which determines not only
the rate but the outcome of dynamical evolution of satellite stellar
systems in the halo of large galaxies.  The total mass of compact
satellites is preserved while they orbit around the outer halo with
cut-off radii smaller than their tidal radii.  These entities
undergo orbital decay and encounter increasingly strong external tidal
perturbation.  The time scale of the orbital evolution is determined
by the magnitude of both the mass and $\ln \Lambda$ which itself
decreases with the slope of the density fall off at outer regions of
the satellite.  Due to the enhanced effect of close encounters, the
magnitude of $\ln \Lambda$ for a cluster with a point mass potential
increases with the numerical resolution.  However, for clusters with a
realistic density distribution, the magnitude of $\ln \Lambda$
converges in the high resolution limit and it can be reliably measured
in the limit of negligible mass loss.

As satellite stellar systems undergo orbital decay, they encounter
increasingly strong tidal perturbation from their host galaxies.
Stars in the outer regions of loosely bound clusters and dwarf
galaxies become detached and form a tidal debris.  But the orbits of
these detached stars do not change abruptly and they diffuse away from
the vicinity of their parent clusters or dwarf galaxies on the orbital
time scales.  Near the tidal radius of their parent clusters and dwarf
galaxies, individual detached stars exchange momentum and energy with
the dark matter halo at comparable rates as the bound stars.  Even
though they are already detached, the escapers continue to maintain
modest gravitational interaction with the bound stars such that they
contribute to the collective dynamical friction effect.

In this paper, we provide a numerical model of a cluster which is
undergoing tidal disruption.  By fitting the simulated orbital decay
with conventional formulae, we show that the residual bound stars alone
cannot fully account for the rapid rate of orbital decay induced by
dynamical friction.  We show that particles which become recently
unbound but remain close to the object also contribute to the effect
of dynamical friction.  In our 'realistic' model, the evidence for
this effect is systematically shown with a considerable amount of
uncertainty during the initial stage of mass loss.  But this general
result is enhanced dramatically during the last phases of the
dissolving process.  Our model star clusters continue to undergo
orbital decay even after all particles have already become unbound.
Due to the fact that our code is able to use millions of particles we
are able to trace the density enhancement of the dissolved star
cluster for quite some time after its dissolution and we found that
this moving cluster of unbound stars continue to undergo orbital
decay.

We propose that the mass responsible for dynamical friction is the sum
of that in the cluster and a large fraction of the mass of the stars
in the tidal debris which follow similar orbits.  The additional mass
to be taken into account are those stars which have the same velocity
and are contained in a ring centred on the cluster's original orbit
and with a half width comparable to the original radius of the parent
cluster.  Our results differ slightly from the findings of
\citet{fuj06} who suggested that only the trailing arm and the unbound
stars in the direct vicinity enhance the dynamical friction.  Our
results are in better agreement with taking both leading and trailing
arm into account, as long as the particles are in a similar orbit as
the remaining satellite.  Still, regarding the large error-bars, our
data-points agree with a simple rule-by-thumb (which has no physical
explanation).  We have shown that one has to take approximately half
of the mass which is lost into account.  This rule-by-thumb is valid
at least as long as we could trace the orbit of our models and is
similar in a quantitatively way to the findings of \citet{fuj06}.
This simple rule-by-thumb remains valid even after the cluster has
become completely tidally disrupted into a moving cluster of unbound
stars.

While the results from this study and from \citet{fuj06} during the
dissolution phase are very similar, we are able to trace the decay of
the satellite even beyond the total destruction, (at which stage
\citet{fuj06} stop their simulations).  We showed for the first time
that a satellite which exists only as a moving cluster of an unbound
density enhancement still suffers dynamical friction and sinks towards
the centre of the host system.

This result has some major implications for recent astronomical
questions.  It increases the possible distance range of the birth
place of the central star cluster in our Milky Way.  In the case of
star clusters sinking to the Galactic Centre during their very short
lifetimes this result could alter the conclusion of a study.
Especially the fact that dissolved clusters continue their orbital
decay at least for a few orbits really enhances the parameter range
from where the central stars were from.

Tidal debris are found around many satellite galaxies in the Galactic
halo.  For example, the mass estimated for the stellar stream
associated with the Sagittarius dwarf spheroidal galaxy is more than
an order of magnitude larger than the total residual mass within that
galaxy \citep[e.g.][]{hel04,law05,fel06}.  The Magellanic stream,
which is mainly composed of neutral hydrogen, also has a mass which is
more than a few percent of that in the Large Magellanic Cloud
\citep[e.g.][]{lin77,lin82,con06}.  There are also suggestions that several
satellite dwarf galaxies lie on various great circles in the sky
\citep{lyn95} perhaps as debris of much larger entities.  In this
context, our results also imply that during their tidal disruption,
the orbital decay of dwarf galaxies in the Galactic halo is sustained
at a rate much larger than that extrapolated from their instantaneous
declining mass.  The contribution of detached stars enhances the
strength of the dynamical friction and promote the completion of the
dwarf galaxies' tidal disruption.  Without it, the orbital decay of
the dwarf galaxies would be halted while many remnant cores would be
preserved.  This process provides a possible scenario to attribute the
``missing satellite'' problem to their enhanced dynamical friction and
effective tidal disruption.

There have been many attempts to reconstruct the orbital evolution of
the dwarf galaxies from their present-day properties.  Our results
suggest that these models may need to be modified to take into account
the efficient dynamical friction, especially during the epoch when
a substantial fraction of their halo dark matter has already become
detached from the satellite galaxies.  Our results also indicate that
the current location where the tidal debris are found may not
necessarily be the location where they became detached from their host
satellite galaxies.

The most and second most massive globular cluster around the Galaxy
are $\omega$-Cen and M22.  The metallicity dispersion within these
cluster is in strong contrast to all other cluster.  But similar
metallicity dispersion has been observed in dwarf spheroidal galaxies
\citep{mcw05}.  It has been suggested that these clusters may be the
cores of disrupted dwarf spheroidal galaxies.  M22 actually resides in
the central region of the Sagittarius dwarf spheroidal galaxy.  The
results we have presented above provide supporting evidence that
dynamical friction is efficient in inducing satellites' orbital decay,
especially during the break-up stage when a large fraction of the
original mass, either in the form of stars in the outer envelope or
loosely found dark matter, become detached from the residual satellite
galaxies.  The enhancing contribution from the break-up stars may
indeed have promoted the disruption of the original parent galaxy and
deposit its nucleus ($\omega$-Cen) at the position where it is found
today. \\

We thank Dr.\ R. Spurzem for useful conversation.  This work was
supported in part by NASA (NAG5-12151) and the California Space
Institute.

\label{lastpage}
\end{document}